\documentclass[aps,pra,preprint,showpacs,superscriptaddress,amsfonts,floatfix]{revtex4}
\usepackage{graphics}

\usepackage[dvips]{graphicx}
\usepackage{amsmath}

\begin{document}

%
%

\title{Swift charged particles in a degenerate electron gas: \\ An estimation for the Barkas effect
in stopping}

%
%

%
\author{I. Nagy}
\affiliation{Department of Theoretical Physics,
Institute of Physics, \\
Budapest University of Technology and Economics, \\ H-1521 Budapest, Hungary}
\affiliation{Donostia International Physics Center, P. Manuel de
Lardizabal 4, \\ E-20018 San Sebasti\'an, Spain}
\author{ I. Aldazabal}
\affiliation{Centro de F\'{i}sica de Materiales (CSIC-UPV/EHU)-MPC,
P. Manuel de Lardizabal 5, \\ E-20018 San Sebasti\'an, Spain}
\affiliation{Donostia International Physics Center, P. Manuel de
Lardizabal 4, \\ E-20018 San Sebasti\'an, Spain}

\date{\today}
\begin{abstract}

Applying a recently suggested new form [Phys. Rev. A {\bf 94}, 042704 (2016)]
for the stopping power in terms of scattering phase shifts, here
we show analytically that an exact leading phase shift may
contain that information which is completely enough to characterize the asymptotic 
charge-sign-independent Bethe term ($\propto{Z_1^2}$) and a
charge-sign-dependent Barkas term ($\propto{Z_1^3}$) in the stopping force of a degenerate 
electron gas for fast projectiles with charge $Z_1$. Our analytic implementation is based
on a Hulth\'en-type potential with velocity-dependent screening.
The next term in an asymptotic expansion, the Bloch term ($\propto{Z_1^4}$), 
measures the difference of the exact and first-order 
Born treatments with Coulomb potential. 
We found a reduced value for the Barkas term, in comparison with the
conventional estimation which rests on the transport cross section.

\end{abstract}


\maketitle

\newpage

\section{Introduction}

The first-order Born approximation for the phase shifts or for the real scattering 
amplitude is a familiar and convenient approximation for handling scattering problems.
It can be adequate, or informative in so many cases that one 
tends to develop the habit of using it without always {\it checking} the conditions 
for its applicability \cite{Peierls79}.
However, when one considers the interaction of an electron via regularized potentials with swift
attractive or repulsive charges moving in a degenerate charged fermion gas,
the experimentally measurable quantity, the stopping power of this gas for projectiles,
may show a deviation from the first-order estimation which results in the Bethe term ($\propto{Z_1^2}$). 
The sign-dependent deviation from this leading term at high velocities is 
characterized by the next-to-leading term, i.e., the Barkas term ($\propto{Z_1^3}$) in the stopping force. 
The existence of such a term in the stopping power of solids is well established experimentally,
in particular at {\it random}-collision condition \cite{Andersen89,Moller97} involving swift
protons and antiprotons ($Z_1=\pm{1}$). 
The next higher-order term to the Bethe and Barkas terms in stopping 
is the so-called Bloch term ($\propto{Z_1^4}$).

The magnitude and interplay of the Barkas and Bloch terms in modulating the leading Bethe term gave
the background to dedicated early experiments \cite{Golovchenko81} 
with bare intruders with atomic number $9\leq{Z_1}\leq{17}$ moving at a
high velocity ($v\simeq{11}v_0$) in {\it channeling} direction of a solid.
From data-analysis it was concluded, using Lindhard's asymptotic theory \cite{Lindhard76},
that either the higher-order terms, i.e., the Barkas and Bloch, are small to detect them or 
both may be large but virtually canceling each other out. However, Peierls' remark \cite{Peierls79}
is relavant to \cite{Golovchenko81}, since the Sommerfeld parameter 
$\gamma=Z_1e^2/(\hbar v)$ is not small. In fact, it is in the range of unity.
Besides, Lindhard clearly emphasized in his work: 
{\it I will try to show that the theory is
quite simple basically, but this does not mean that I am sure of all details of it}.

Notice at this point, that a more recent mean-field calculation \cite{Borisov07}, performed in 
the auxiliary orbital representation of
time-dependent density-functional theory with $Z_1=\pm{1}$ and $(v/v_0)\in{(0,6)}$
for metallic clusters,
results in a very small Barkas effect already for about $v>3v_0$. 
The above, experimental and theoretical, facts give the motivations to our contribution. 
We implement, {\it analytically},
a recently proposed form \cite{Grande15} for the stopping power
which is also given in terms of scattering phase shifts. That form was termed as superior to the
one used in \cite{Lindhard76}. Moreover, based on a numerical implementation, its possible
relevance to understanding the charge-sign-dependent term was explicitly stressed \cite{Grande15}.
  
This paper is organized as follows. The next Section is devoted to the theory and the discussion
of the results obtained. The last Section contains the summary, and our comments.
We will use natural (rather than Hartree atomic, where $e^2=\hbar=m=1$) units 
in this work, except where (in an illustrative Figure) the opposite is explicitly stated.


\section{Theory and results}

In energy-loss experiments we add an external, heavy charged projectile to the many-body
system of {\it interacting} electrons. In a scattering description, the incoming one-electron states
are plane waves due to the translational invariance of the target system with number density $n_0$. 
The momentum distribution of these states of {\it real} electrons is described by a 
renormalized distribution function. We are, therefore, at an important exceptional case where
Landau's description for quasiparticle energies with a noninteracting-like Fermi-Dirac
distribution function is not applicable \cite{Migdal77}. Similarly, the simpler 
Kohn-Sham-like mean-field modeling
with an ideal (step-function) distribution function to consider an averaging over 
momentum distribution, is incomplete. This problem is not investigated yet at low \cite{Puska83,Nagy89} 
and intermediate \cite{Salin99} velocities of charged projectiles. Peliminary
results for the case of a fixed charged impurity are available in \cite{Nagy17}.
However, at high projectile velocities ($v$), the relative velocity ($v_r$) to the underlying
two-body kinematics in stopping is given by
\begin{equation}
v_r\, =\, v\left[1+\frac{1}{3}\, \frac{\langle{v_e^2}\rangle}{v^2}\right] \nonumber
\end{equation}
where $v_e$ is the electron velocity \cite{Brandt81}. In an interacting system $v_e\in(0,\infty)$, even
at zero temperature. But, in spite of $\langle{v_e^2}\rangle >[(3/5)(3\pi^2 \hbar^3 n_0/m)^{2/3}]$,
at high velocities we can take, as expected, $v_r=v$. 
Thus, the wave number of an electron becomes $k=mv/\hbar$.

In order to motivate, heuristically as a first step, the basic expression used in this work
for asymptotic stopping power calculation, we start by the Coulomb-potential case. Taking
the Coulomb phase shifts \cite{Messiah61} in a partial-wave expansion, we can write
\begin{equation}
\sum_{l=0}^{\infty}(l+1)\sin^2[\sigma_l(k)-\sigma_{l+1}(k)]\equiv{\gamma\,
\sum_{l=0}^{\infty}
\sin[\sigma_l(k)-\sigma_{l+1}(k)]\cos[\sigma_l(k)-\sigma_{l+1}(k)]} 
\end{equation}
where $\gamma=Z_1 e^2/{\hbar v}=Z_1/(a_0 k)$, is the Sommerfeld parameter.
The common Coulomb logarithm \cite{Messiah61} does not appear
in the differences $[\sigma_l(k)-\sigma_{l+1}(k)]=\arctan[\gamma/(l+1)]$. However,
in the case of a screened (regularized) potential, $V(r)$, we get an inequality
\begin{equation}
\sum_{l=0}^{\infty}(l+1)\sin^2[\delta_l(k)-\delta_{l+1}(k)]\neq{\gamma\,
\sum_{l=0}^{\infty}
\sin[\delta_l(k)-\delta_{l+1}(k)] \cos[\delta_l(k)-\delta_{l+1}(k)]       } 
\end{equation}
where the Bessel phase shifts are denoted, as ususal, by $\delta_l(k)$. 
With a screened potential both sides of this inequality are finite and
the left-hand-side is positive definite always. A truncated summation in the equality of Eq.(1),
results in
\begin{equation}
\sum_{l=0}^{L_{max}}\, (l+1)\sin^2[\sigma_l(k)-\sigma_{l+1}(k)]=\gamma^2\, \sum_{l=0}^{L_{max}}
\frac{l+1}{(l+1)^2+\gamma^2}. \nonumber 
\end{equation}
With $L_{max}\rightarrow{\infty}$ one would get a divergency, as it is well-known with a
three-dimensional Rutherford differential cross section to angle-weighting in the transport cross section.

Application
of the left-hand-side of Eq.(2) is common \cite{Lindhard76,Nagy89} in transport 
cross section calculations with
spherical scattering potentials. The right-hand-side corresponds, {\it precisely}, to the
new form proposed recently and considered as a superior form \cite{Grande15}. That
proposal is based on calculating a noncentral induced electron density from which
the stopping force is defined by integrating over the gradient of the corresponding (via convolution)
electrostatic (Hartree) potential. This definition for the retarding force is 
classical in the sense that it does not reflect the probabilistic interpretation of 
quantum mechanics via complex matrix elements of a force operator. Indeed, the left-hand-side
of Eq.(2) is expressible \cite{Gyorffy72,Suhl75,Bonig89,Tang98,Zawadowski09} as 
\begin{equation}
\sum_{l=0}^{\infty}\, (l+1) \left|i^l(-i)^{l+1}\, e^{i[\delta_l(k)-\delta_{l+1}(k)]}\,
\left(\int_0^{\infty}dr r^2 R_l(r)
\frac{\partial{U(r)}}{\partial{r}} R_{l+1}(r)\right)\right|^2,  \nonumber
\end{equation}
where, for simplicity, the $k$-dependence is not explicit in $R_l(k,r)$
which is a radial component in the partial-wave-based approach 
with spherical $U(r)=(2m/\hbar^2)V(r)$.
Contrary to this, the right-hand-side of Eq.(2) corresponds, {\it formally}, to the real 
part of complex elements
\begin{equation}
\sum_{l=0}^{\infty}\, (l+1) Re \left\{i^l(-i)^{l+1}\, e^{i[\delta_l(k)-\delta_{l+1}(k)]}
\left[\int_0^{\infty}dr r^2 R_l(r)\frac{2m}{\hbar^2}
\frac{\partial}{\partial r}\left(-\frac{Z_1 e^2}{r}\right) R_{l+1}(r)\right]\right\}, \nonumber
\end{equation}
which has a simple mixed-product representation with two real terms as well
\begin{equation}
\sum_{l=0}^{\infty}\, (l+1) 
\left(\int_0^{\infty}dr r^2 R_l(r)
\frac{\partial{U(r)}}{\partial{r}} R_{l+1}(r)\right)
\left(\int_0^{\infty}dr r^2 R_l(r)\frac{2m}{\hbar^2}
\frac{Z_1 e^2}{r^2} R_{l+1}(r)\right). \nonumber
\end{equation}
The structure of this product shows that for swift projectiles, where
the radial functions deviate only slightly from the corresponding plane-wave components, 
it will result in larger values than the square
\cite{Gyorffy72,Suhl75,Bonig89,Tang98,Zawadowski09} of its first term  
with consistently employed $\partial U(r)/\partial r$.

In this paper we will implement the right-hand-side of Eq.(2), and apply it 
at high velocities of projectiles to the recently proposed
\cite{Grande15} stopping power expression
\begin{equation}
\frac{dE}{dz}\, =\, (mv^2)\, n_0\, \frac{2\pi}{k^2}\, \gamma\, \sum_{l=0}^{\infty}\, 
\sin\{2[\delta_l(k)-\delta_{l+1}(k)]\}.
\end{equation}
Since we focus on the $|\gamma|<1$ case in this work, we apply a two-term Taylor expansion
\begin{equation}
\sum_{l=0}^{\infty}\,
\sin\{2[\delta_l(k)-\delta_{l+1}(k)]\}\simeq{2\delta_0(k) - \frac{4}{3}\,
\sum_{l=0}^{\infty}[\delta_l(k)-\delta_{l+1}]^3}. 
\end{equation}
The dominant role of the leading ($l=0$) phase shift is transparent
in this asymptotic form. We proceed
by using a Hulth\'en-type \cite{Hulthen42} effective scattering potential
\begin{equation}
V(r)\, =\, -\, \frac{Z_1 e^2 \Lambda}{e^{\Lambda r} -1},
\end{equation}
with $Z_1\geq{1}$ for attractive (bare) intruders. For antiprotons $Z_1=-1$. The
screening parameter $\Lambda$ will be fixed below. For this potential,
one has an exact solution
for the Jost function, $F_0(k)$, of scattering theory \cite{Jost47,Jost51,Weinberg63}. The product
representation for this function
\begin{equation}
F_0(k)\, =\, \prod_{n=1}^{\infty} \left[1 + \frac{i (Z_1/a_0)}{n (k -i n\Lambda/2)}\right], \nonumber
\end{equation}
allows the determination of the leading phase shift 
from $e^{2i\delta_0(k)}=F_0(k)/F_0(-k)$. We get
\begin{equation}
\delta_0(k)=\sum_{n=1}^{\infty}\arctan\left\{\frac{\gamma}{n 
[1-(\gamma \Lambda/2k) +(n\Lambda/2k)^2]}\right\}.
\end{equation}
This is an {\it exact} result. 
With $(\Lambda/2k)\ll{1}$ in Eq.(6) one has
\begin{equation}
\sum_{n=0}^{N_{max}}\, \arctan\frac{\gamma}{n+1}\, =\, 
{\sum_{l=0}^{L_{max}}[\sigma_l(k)-\sigma_{l+1}(k)]}. \nonumber  
\end{equation}
Therefore, an approximation \cite{Golovchenko81} for $L_{max}$ must respect the
$(L_{max}\Lambda/2k)\ll{1}$ constraint.

Next, we determine the first-order Born result for $\delta_l^{B}(k)$ from
\begin{equation}
\delta_l^{B}(k)\equiv{-2k\, \frac{m}{\hbar^2}\int_{0}^{\infty}dr\, r^2\, V(r)\, j_l^2(kr)}=
-\frac{m}{\hbar^2}\frac{1}{4\pi} \frac{1}{k}\int_0^{2k}dq\, q\, V(q)\, P_l(x), 
\end{equation}
where $j_l(kr)$ is a spherical Bessel function of the first kind and $V(q)$ 
is the Fourier transform of a spherical potential $V(r)$. To the
Legendre polynomials, we have $x=1-q^2/(2k^2)$ as argumentum.
In the Hulth\'en case, at $\Lambda\neq{0}$, we can write
\begin{equation}
V(r)\, =\, -Z_1e^2\Lambda\sum_{n=1}^{\infty} e^{-n\Lambda r} \nonumber
\end{equation}
\begin{equation}
V(q)\, =\, 8\pi Z_1e^2 \Lambda^2\, \sum_{n=1}^{\infty}\frac{n}{[q^2 +(n\Lambda)^2]^2} \nonumber
\end{equation}
which make the calculation for the important $\delta_0^{B}(k)$ easy. In that case $P_0(u)=1$
in Eq.(7), and thus we obtain in first-order (plane-wave-based) Born approximation
\begin{equation}
\delta_0^{B}(k)\, =\, \sum_{n=1}^{\infty} 
\frac{\gamma}{n[1 +(n\Lambda/2k)^2]}\,
=\, \frac{Z_1}{a_0 k}\, [Re\, \psi(1+i2k/\Lambda) -\psi(1)],
\end{equation}
in terms of standard digamma functions. In our asymptotic ($u\gg{1}$) case, we can apply the 
$[Re\, \psi(1+iu) -\psi(1)]=(1/2)\ln(1+\Gamma^2 u^2]$ approximation \cite{Ferrariis84},
where $\Gamma=1.781$. Comparison of Eq.(6) and Eq.(8) shows the fine {\it details}
behind the applicability \cite{Peierls79} of the
first-order Born approximation. The phase shift must be small (i.e., $|\gamma|<1$)
and, in addition, one has to neglect
the charge-sign-sensitive ($\propto{\gamma}$) term in the denominator of Eq.(6).


Now, let us go back to Eq.(2) and evaluate both sides with first-order Born approximation 
for the real scattering amplitude by considering the right-hand-side ($rhs$) of Eq.(7) as well.
From that evaluation, we get the following, very transparent, inequality
\begin{equation}
\int_{0}^{2k} dq\, q^3\, [V(q)]^2\, < \, \int_{0}^{2k} dq\, q^3 \, 
\left[\frac{4\pi Z_1 e^2}{q^2}\, V(q)\right]. \nonumber
\end{equation}
The $lhs$, which rests on the conventional \cite{Lindhard76} transport cross section, 
tends more {\it slowly} (from below) to the Bethe limit than the $rhs$ 
which is based on the new form in \cite{Grande15}.

At this point, we return to the prescription of a reasonable screening parameter $\Lambda$.
In order to follow as close as possible the motivating work in \cite{Grande15}, where a Yukawa-type
effective potential, $V(r)=(-Z_1 e^2/r)\exp(-\lambda r)$, was used to numerics, 
we employ a scaling argument. Namely, we require the
equality of the corresponding leading phase shifts at the Born limit. This constraint
results in $\Lambda=1.781\lambda$. Remarkably enough, this scaling is in a very nice agreement
with an earlier \cite{Dutt85} one $\Lambda=1.75\lambda$. This latter
is based on an entirely different, bound-state-related, physical problem.
Following \cite{Nagy96,Arista98} we take, as in the motivating \cite{Grande15}, the 
$\lambda=(\omega_p/v)$ prescription, where $\omega_p=(4\pi n_0 e^2/m)^{1/2}$
is the classical plasma frequency.

With a two-term Taylor expansion, $\arctan(\alpha)\simeq{[\alpha-(1/3)\alpha^3]}$ in Eq.(6), 
and considering Eqs.(3-4) and Eqs.(7-8) as well, we reduce to the 
traditional \cite{Lindhard76} asymptotic form
\begin{equation}
\frac{dE}{dz}\, =\, m (\omega_p^2\, a_0)\left(\frac{Z_1}{a_0 k}\right)^2
\left[L_0 +Z_1 L_1 + Z_1^2 L_2 \right],
\end{equation}
with Bethe ($L_0)$, Barkas ($L_1$), and Bloch ($L_2$) terms, the following expressions 
\begin{equation}
L_0\, =\, \frac{1}{2}\ln\left[1+\left(\frac{2k\Gamma}{\Lambda}\right)^2\right]
\simeq{\ln\frac{2mv^2}{\hbar\omega_p}},
\end{equation}
\begin{equation}
L_1=\frac{\Lambda}{2a_0 k^2}\, L_0\simeq{0.89\, \frac{(\omega_p/v)}{a_0 k^2} 
\ln\frac{2mv^2}{\hbar\omega_p}}=0.89\, \frac{e^2\omega_p}{m v^3}\,
\ln\frac{2mv^2}{\hbar\omega_p}, 
\end{equation}
\begin{equation}
L_2\, =\, -\frac{1}{3}\left(\frac{1}{a_0 k}\right)^2\, \zeta(3)\, -
\frac{2}{3}\left(\frac{1}{a_0 k}\right)^2\, \zeta(3) \simeq
{-1.2\,\left(\frac{v_0}{v}\right)^2}. 
\end{equation}
In this equation the (1/3)-part comes from the second term of the above expansion for
$\arctan(\alpha)$ in the {\it leading} phase shift, and the (2/3)-part rests on the 
second term in Eq.(4). Both terms are
calculated for the $\Lambda\rightarrow{0}$ case. This is an allowed approximation
in the asymptotic limit since from Eq.(7) one can get easily (with a Yukawa-form) the estimation
\begin{equation}
[\delta_l^{B}(k)-\delta_{l+1}^{B}(k)]\simeq{\frac{\gamma}{(l+1)}\, [1-O(\lambda^2/k^2)]}. \nonumber
\end{equation}
Therefore, the Bloch term with screening tends from {\it below} to its Coulomb equivalent.

From the pioneering work \cite{Lindhard76} of Lindhard ($L$) one has
to the Barkas ($\propto{Z_1^3}$) term 
\begin{equation}
L_1^{(L)}\, =\, \beta\, \frac{e^2\omega_p}{m v^3}\, L_0\, =\, 
\beta\, \frac{e^2\omega_p}{m v^3}\, \ln\frac{2mv^2}{\hbar\omega_p},
\end{equation}
where $\beta=\pi$ or $\beta=(3/2)\pi$. 
One can see that this is essentially {\it larger} than ours in Eq.(11).
Our result is in nice harmony
with a previous one \cite{Borisov07} obtained
in a self-consistent, orbital-based mean-field approximation within TDDFT
with cluster targets and $Z_1=\pm{1}$ for projectiles. In that numerical modeling, 
the Barkas effect is also very small for $(v/v_0)>3$.

Lindhard's insightful estimation is based on a simple
shifting, as $[k^2 -(Z_1\beta\lambda/a_0)]$, in the scattering energy and using that shifted
quantity in the pre-factor of a leading Bethe term. However, his leading term was
calculated using the transport cross section, i.e., the $lhs$ of the above inequality with real
scattering amplitudes. It is easy to show, by taking a Yukawa-type modeling, that the
mentioned inequality results in asymptotically
\begin{equation}
\left(\ln\frac{2mv^2}{\hbar\omega_p}-\frac{1}{2}\right)\, <\, \ln\frac{2mv^2}{\hbar\omega_p}. \nonumber
\end{equation}
We must note at this point, in favor of a late expert, that this $lhs$ together with
Lindhard's estimation for the Barkas term, would result in a reasonable numerical agreement with our
present result, which is based on a different \cite{Grande15} modeling, for the case of proton $Z_1=1$
due to certain cancelation. But, in the case of an antiproton $Z_1=-1$, it would give  
a much stronger deviation from the Bethe limit.

After the above detailed analysis related to the stopping power of an electron gas for
swift-projectiles, we continue with a quantitative comparison at 
high-velocities. We use to this comparison theoretical
results from an {\it independent} source. Thus, in Fig.1, we exhibit
our results for proton and antiproton, by solid and dashed curves, respectively.
The discrete dots on the dotted curve are  taken from \cite{Correa15} for $Z_1=1$. They are based
on large-scale numerics \cite{Correa15} performed within TDDFT considering the valence part of a 
free-electron-like solid, Al. One can see a nice harmony between the solid curve and dots.

In a more general content, we should note that
in the folklore of numerics \cite{Borisov07,Correa15} it is customary
to employ the first-principles and benchmark wording
in order to emphasize a method-capability {\it a priori}. However, that wording is true only 
with a small but important modification. 
TDDFT, as well as the time-independent DFT, are first-principles 
approaches with semi-empirical inputs \cite{Cohen08}. 
Thus they have an {\it a posteriori} character, as was emphasized in a pioneering work \cite{Puska83}. 
Comparisons with high-precision data
or with exact \cite{Inigo18} solutions on correlated models are always required
in order to establish a transferable knowledge.

\begin{figure}
\scalebox{0.4}[0.4] {\includegraphics{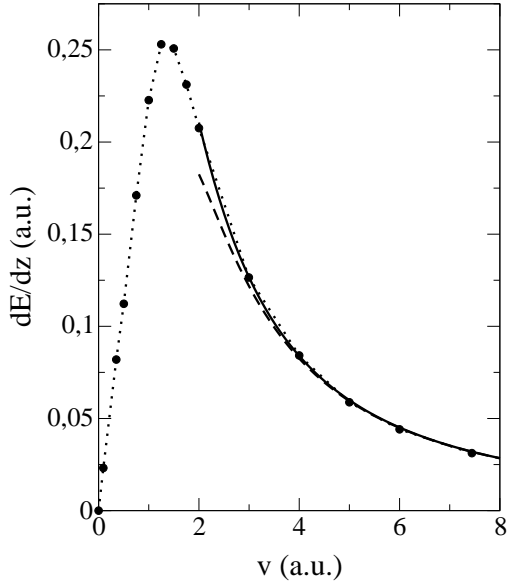}}
\caption{Stopping powers, in atomic units, of an electron gas with $r_s=2.07$ for 
projectiles with $Z_1=\pm{1}$. The solid ($Z_1=1$) and dashed ($Z_1=-1$) curves refer to the
analytical approach detailed in this contribution. The numerics-based discrete 
dots (for $Z_1=1$) are taken from \cite{Correa15}. 
\label{figure1}}
\end{figure}
%


The two sides of the inequality, in Eq.(2), represent two approximations which give
concordant results, mathematically, only for the $\Lambda=0$ case at which they yield
a divergent expresssion for an observable quantity. The $lhs$ of Eq.(2) seems to be \cite{Salin99}
the correct one at $v\rightarrow{0}$, where the relative wave number is determined
by the Fermi velocity. In that case the probabilistic interpretation
of quantum mechanics, with squares of force matrix-elements, becomes transparent.
Beyond that
impurity limit and in a mean-field picture, which rests on projectile screening and
scattering of independent electrons, a {\it nonspherical} multipole expansion \cite{Hill67} could be the
consistent  one. However, in such a treatment with nonspherical scattering potentials to determine the
corresponding force-matrix elements \cite{Suhl75}, one arrives at coupled equations. A detailed investigation 
of this quantum-mechanical, probabilistic, approach is out of the scope of the present work and
is left to a dedicated study.

The mathematical divergency of summations in Eq.(1) can be cured by taking a
regularized (screened) two-body interaction. But that step can result in two different
{\it interpretations} as well, as our diagnosis after Eq.(2) shows. We believe,
therefore, that the two-dimensional version of the stopping problem \cite{Nagy95,Zaremba05} 
may help
in such interpretation-difficulties. There, with Coulomb interaction energy, we get
to the corresponding transport cross section
\begin{equation}
\sum_{m=0}^{\infty}\sin^2[\sigma_m(k)-\sigma_{m+1}(k)]\equiv{
\sum_{m=0}^{\infty}\frac{\gamma}{(m+1/2)}\sin[\sigma_m(k)-\sigma_{m+1}(k)]
\cos[\sigma_m(k)-\sigma_{m+1}(k)]}, \nonumber
\end{equation}
using the fact that $[\sigma_m(k)-\sigma_{m+1}(k)]=\arctan[\gamma/(m+1/2)]$. 
The summation \cite{Nagy95} in the above equation ($k=mv/\hbar$) results in a non-divergent expression
for the stopping power
\begin{equation}
\frac{dE}{dx}=n_0(mv^2)\frac{\gamma}{k} 2\pi \tanh(\pi\gamma)
=n_0(mv^2) \frac{Z_1 e^2}{mv^2} 2{\pi}
\tanh\left(\frac{\pi Z_1 e^2}{\hbar v}\right).\nonumber
\end{equation}
How a regularization (screening) will change
this exact quantum-result by using the above two sides with the $\sigma_m(k)\rightarrow{\delta_m(k)}$
substitution is a challenging question. 
Answers on it may contribute to our undertanding of interpretation-difficulties as well.


\section{summary and comment}

In this paper we have investigated the problem of the asymptotic behavior
of the stopping force of a degenerate electron gas for fast charged
projectiles. The applied method rests on scattering phase shifts in
order to implement a recently proposed \cite{Grande15} new
form. A Hulth\'en-type potential, with velocity-dependent screening,
is used in order to derive an {\it exact} form for the leading phase shift. 
For charge-conjugated stable particles, i.e., for protons and antiprotons ($Z_1=\pm{1}$),
a reduced value for the sign-dependent Barkas term is deduced. This small
value is in harmony with a previous result \cite{Borisov07} obtained within the  
self-consistent mean-field framework of time-dependent density-functional theory (TDDFT)
applied to cluster targets.
Very reasonable agreement with results \cite{Correa15} based on large-scale numerics in TDDFT
with $Z_1=1$ for an Al target, 
is also established. However, it would be informative to see the corresponding
prediction of such a numerical calculation for antiprotons as well. Indeed, for repulsive 
projectiles one-electron bound states can not pose an additional problem. 

From the above agreements, one might conclude
that the application of the mean-field concept of dynamical {\it projectile} screening
is a reasonable one in order to consider the stopping power of a degenerate fermion
system for swift charged projectiles. 
Our result obtained for the Bethe logarithm, $L_0=\ln(2mv^2/\hbar\omega_p)$,
takes its conventional form. However, as it is well-known \cite{Golovchenko81,Nagy01}, the same
asymptotic leading term comes from Pines' description of 
an {\it interacting} three-dimensional electron gas \cite{Pines63}.
There, due to a canonical transformation treatment on the many-body Hamiltonian, an individual
{\it electron} is screened spherically in two-body scattering and the collective mode represents 
a separate degree in dissipation processes. Because of such screening the sudden scattering with the
swift bare projectile will result in a finite cross section and associated energy transfer.


Straightforward application \cite{Nagy01} of Pines' modeling results in a Barkas
term $L_1\propto{\Lambda/v^2}$ with $\Lambda=1.781\lambda_{TF}$, where the static Thomas-Fermi ($TF$)
parameter is about $\lambda_{TF}\simeq{1/\sqrt{r_s}}$. In such a two-channel modeling, $L_1$ and $L_2$ have
similar ($\propto{v^{-2}}$) scaling in the projectile velocity. 
Moreover, such an $L_1\propto{v^{-2}}$ dependence \cite{Nagy01} would fit
to the experiment-based statement made at {\it random} collisional situation \cite{Andersen89}
of stopping measurement with $Z_1=\pm{1}$. It could allow a stronger cancelation between $L_1$
and $L_2$ at {\it positive} $\gamma$ values \cite{Golovchenko81} as well.
Further detailed studies, by considering the sudden-character of a swift projectile in its
time-dependent interaction with localized target-states of solids, are thus desirable \cite{Bruneval18}. 
The observable quantity, the energy loss of a heavy projectile, is well-defined classically, but
the transferable knowledge on it requires treatments at the level of quantum mechanics.

\newpage

\begin{acknowledgments} The authors are thankful to Professor
Alfredo Correa for several, very useful discussions. 
This work was supported partly
by the Spanish Ministry of Economy and Competitiveness (MINECO: Project FIS2016-76617-P).
\end{acknowledgments}


\begin{thebibliography}{00}
\bibitem{Peierls79}
R. Peierls, {\it Surprises in Theoretical Physics} (Princeton University Press, Princeton,
New Jersey, 1979), pp. 3-6.
%
\bibitem{Andersen89}
L. H. Andersen, P. Hvelplund, H. Knudsen, S. P. Moller, J. O. P. Pedersen, E. Uggerhoj,
K. Elsener, and E. Morenzoni, Phys. Rev. Lett. {\bf 62}, 1731 (1989).
%
\bibitem{Moller97}
S. P. Moller, E. Uggerhoj, H. Blume, H. Knudsen, U. Mikkelsen, K. Paludan, and
E. Morenzoni, Phys. Rev. A {\bf 56}, 2930 (1997).
%
\bibitem{Golovchenko81}
J. A. Golovchenko, A. N. Goland, J. S. Rosner, C. E. Thorn, H. E. Wegner,
H. Knudsen, and C. D. Moak, Phys. Rev. B {\bf 23}, 957 (1981).
%
\bibitem{Lindhard76}
J. Lindhard, Nucl. Instrum. Methods {\bf 132}, 1 (1976).
%
\bibitem{Borisov07}
M. Quijada, A. G. Borisov, I. Nagy, R. Diez Mui\~{n}o, and P. M. Echenique,
Phys. Rev. A {\bf 75}, 042902 (2007).
%
\bibitem{Grande15}
P. L. Grande, Phys. Rev. A {\bf 94}, 042704 (2016).
%
\bibitem{Migdal77}
A. B. Migdal, {\it Qualitative Methods in Quantum Theory} (Benjamin, London, 1977), p. 302.
%
\bibitem{Puska83}
M. J. Puska and R. M. Nieminen, Phys. Rev. B {\bf 27}, 6121 (1983).
%
\bibitem{Nagy89}
I. Nagy, A. Arnau, P. M. Echenique, and E. Zaremba, Phys. Rev. B {\bf 40}, R11983 (1989).
%
\bibitem{Salin99}
A. Salin, A. Arnau, P. M. Echenique, and E. Zaremba, Phys. Rev. B {\bf 59}, 2537 (1999).
%
\bibitem{Nagy17}
I. Nagy and M. L. Glasser, in {\it Many-body Approaches at Different Scales},
Editors: G. G. N. Angilella and C. Amovilli, (Springer, New York, 2018), pp. 133-138.
%
\bibitem{Brandt81}
S. Kreussler, C. Varelas, and W. Brandt, Phys. Rev. B {\bf 23}, 82 (1981).
%
\bibitem{Messiah61}
A. Messiah, {\it Quantum Mechanics} (Elsevier, Amsterdam, 1961).
%
%
\bibitem{Gyorffy72}
G. D. Gaspari and B. Gy\"orffy, Phys. Rev. Lett. {\bf 28}, 801 (1972).
%
\bibitem{Suhl75}
E. G. d'Agliano, P. Kumar, W. Schaich, and H. Suhl, Phys. Rev. B {\bf 11}, 2122 (1975).
%
\bibitem{Bonig89}
L. B\"onig and K. Sch\"onhammer, Phys. Rev. B {\bf 39}, 7413 (1989).
%
\bibitem{Tang98}
J.-M. Tang and D. J. Thouless, Phys. Rev. B {\bf 58}, 14179 (1998).
%
\bibitem{Zawadowski09}
I. Nagy and A. Zawadowski, J. Phys.: Condens. Matter {\bf 21}, 175701 (2009).
%
\bibitem{Hulthen42}
L. Hulth\'en, Arkiv Mat. Fys. Astron. {\bf 28A}, no. 5 (1942).
%
\bibitem{Jost47}
R. Jost, Helv. Physica Acta {\bf 20}, 256 (1947).
%
\bibitem{Jost51}
R. Jost and A. Pais, Phys. Rev. {\bf 82}, 840 (1951).
%
\bibitem{Weinberg63}
S. Weinberg, Phys. Rev. {\bf 131}, 440 (1963).
%
\bibitem{Ferrariis84}
L. de Ferrariis and N. R. Arista, Phys. Rev. A {\bf 29}, 2145 (1984).
%
\bibitem{Dutt85}
R. Dutt, K. Chowdhury, and Y. P. Varshni, J. Phys. A: Math. Gen. {\bf 18}, 1379 (1985).
%
\bibitem{Nagy96}
I. Nagy and A. Bergara, Nucl. Instrum. Methods B {\bf 115}, 68 (1996). 
%
\bibitem{Arista98}
A. F. Lifshitz and N. R. Arista, Phys. Rev. A {\bf 57}, 200 (1998).
%
\bibitem{Correa15}
A. Schleife, Y. Kanai, and A. A. Correa, Phys. Rev. B {\bf 91}, 014306 (2015).
%
\bibitem{Cohen08}
A. J. Cohen, P. Mori-S\'anchez, and W. Yang, Science {\bf 321} 792 (2008).
%
\bibitem{Inigo18}
I. Nagy and I. Aldazabal, arXiv: 1810.12796 [quant-ph].
%
\bibitem{Hill67}
A. D. Boardman, A. D. Hill, and S. Sampanthar, Phys. Rev. {\bf 160}, 472 (1967).
%
\bibitem{Nagy95}
I. Nagy, Phys. Rev. B {\bf 51}, 77 (1995).
%
\bibitem{Zaremba05}
E. Zaremba, I. Nagy, and P. M. Echenique, Phys. Rev. B {\bf 71}, 125323 (2005).
%
\bibitem{Nagy01}
I. Nagy, Phys. Rev. A {\bf 65}, 014901 (2001).
%
\bibitem{Pines63}
D. Pines, {\it Elementary Excitations in Solids} (Benjamin, New York, 1963).
%
\bibitem{Bruneval18}
I. Maliyov, J. P. Crocombette, and F. Bruneval, Eur. Phys. J. B {\bf 91}, 172 (2018).
%



\end{thebibliography}
\end{document}